\documentclass[namedreferences]{solarphysics}

\usepackage[hyperref,optionalrh]{spr-sola-addons} 
\usepackage{graphicx}        
\usepackage{color}           
\usepackage{breakurl}        
\usepackage{xcolor}

\begin{document}

\begin{article}
\begin{opening}

\title{Parametrization of sunspot groups based on machine learning approach}

\author[addressref={aff1,aff2},email={egor.illarionov@math.msu.ru}]{\inits{E.A.}\fnm{Egor}~\lnm{Illarionov}\orcid{0000-0002-2858-9625}}
\author[addressref=aff3,email={tlatov@mail.ru}]{\inits{A.G.}\fnm{Andrey}~\lnm{Tlatov}\orcid{0000-0003-1545-2125}}
\address[id=aff1]{Moscow State University, Moscow, Russia}
\address[id=aff2]{Moscow Center of Fundamental and Applied Mathematics, Moscow, Russia}
\address[id=aff3]{Kislovodsk Mountain Astronomical Station, Kislovodsk, Russia}

\runningauthor{Illarionov \& Tlatov}
\runningtitle{\textit{Solar Physics} Parametrization of sunspot groups}

\begin{abstract}
Sunspot groups observed in white-light appear as complex structures. Analysis of these structures is usually based on simple morphological descriptors which capture only generic properties and miss information about fine details. We present a machine learning approach to introduce a complete yet compact description of sunspot groups. The idea is to map sunspot group images into an appropriate lower-dimensional (latent) space. We apply a combination of Variational Autoencoder and Principal Component Analysis to obtain a set of 285 latent descriptors. We demonstrate that the standard descriptors are embedded into the latent ones. Thus, latent features can be considered as an extended description of sunspot groups and, in our opinion, can expand the possibilities for the research on sunspot groups. In particular, we demonstrate an application for estimation of the sunspot group complexity. The proposed parametrization model is generic and can be applied to investigation of other traces of solar activity observed in various spectrum lines. Key components of this work, which are the parametrization model, dataset of sunspot groups and latent vectors, are available in the public GitHub repository \url{https://github.com/observethesun/sunspot_groups} and can be used to reproduce the results and for further research.

\end{abstract}
\end{opening}

\section{Introduction}

Sunspots appear as prominent features in the solar photosphere. Comparing white-light solar disk images and magnetograms one can note that sunspots are also associated with strong magnetic fields. In our paper we will focus only on white-light solar disk images due to the primary interest in historical observations which miss magnetic field information.

Typically, sunspots are not individual features but appear in groups. It was noted that for estimation of solar activity the number of groups is as important as the number of sunspots \citep{Hoyt}. In many cases an attribution of sunspots to sunspot groups is not a trivial problem and includes more aspects than just closeness between sunspots. In our research we will analyze sunspot groups formed by expert observers and thus omit this problem. However, it should be noted that there were several attempts to automate this step (e.g. \citet{Colak}, \citet{Zharkova}). 

What we focus on is how to describe the structure of sunspot groups numerically and to preserve as much information as possible. It is typical to describe the structure of a sunspot group by its area, elongation, number of spots and some other simple morphological properties. Probably, the most advanced and informative descriptor is given by the class of sunspot group according to Zurich or Modified Zurich classification systems \citep{McIntosh}. Indeed, the proposed sunspot group class encodes together area, elongation and structural information about the group.

Obviously, the above set of descriptors is far from being complete in a sense that it does not allow detailed reconstruction of sunspot group structure. At the first glance it is also unclear how to extend this set substantially in a computationally feasible way.
There were several attempts to introduce additional morphological descriptors, see e.g. \citet{Stenning},  \citet{Ternullo} and \citet{Makarenko}.
However, the problem of automated complete sunspot group description remains actual.

In our research we propose a data-driven approach to sunspot group parametrization which provides a compact yet complete set of sunspot group descriptors. The idea is to map sunspot group images into an appropriate lower-dimensional space. We will construct this mapping by training the Variational Autoencoder model and selecting the features using the Principal Components Analysis. While autoencoder models based on neural networks were already used in some papers e.g. by \citet{Chen} to obtain useful features for solar flare prediction and by \citet{Sadykov} for spectrum lines compression, however, we first provide a systematic analysis of the latent space and demonstrate that at least several latent features have a clear physical interpretation.
   
\section{Data}

We use a daily catalogue of sunspot group images provided by the Kislovodsk Mountain Astronomical Station for the period 2010--2020 \citep{Kislovodsk}. Initially, the solar disk was observed at the station in white-light and photoheliograms were recorded. The photoheliograms were processed to isolate sunspots, sunspot cores and pores and attribute them to sunspot groups. Each step of this process is verified by expert observers and thus we consider this data as ground truth in the present research. The resulting daily sunspot maps and group attribution are available on the website \url{https://observethesun.com}.

To prepare a dataset of sunspot groups, we rescale all sunspot maps to the constant solar disk radius equal to 1200 pixels. The constant value is selected such that most sunspot groups can be contained in patches of size $256\times 256$ pixels. Then we crop patches of size $256\times 256$ pixels centered at each sunspot group  that is completely within the patch frame. To be more detailed, we obtain 3 binary masks of size $256\times 256$, one for sunspots, cores and pores. Note that by construction binary masks are constructed for a particular group only and ignore any information about other groups. To avoid undesired projection effects near the solar limb, we consider only groups with central meridian distance below $60^{\circ}$. There are 8498 groups following this condition.

The constructed dataset of sunspot groups is publicly available in the GitHub repository \url{https://github.com/observethesun/sunspot_groups}. Note that we also provide meta information about the groups including its area, location, number of spots and elongation. 

\begin{figure}
\centering
\includegraphics[width=0.35\textwidth]{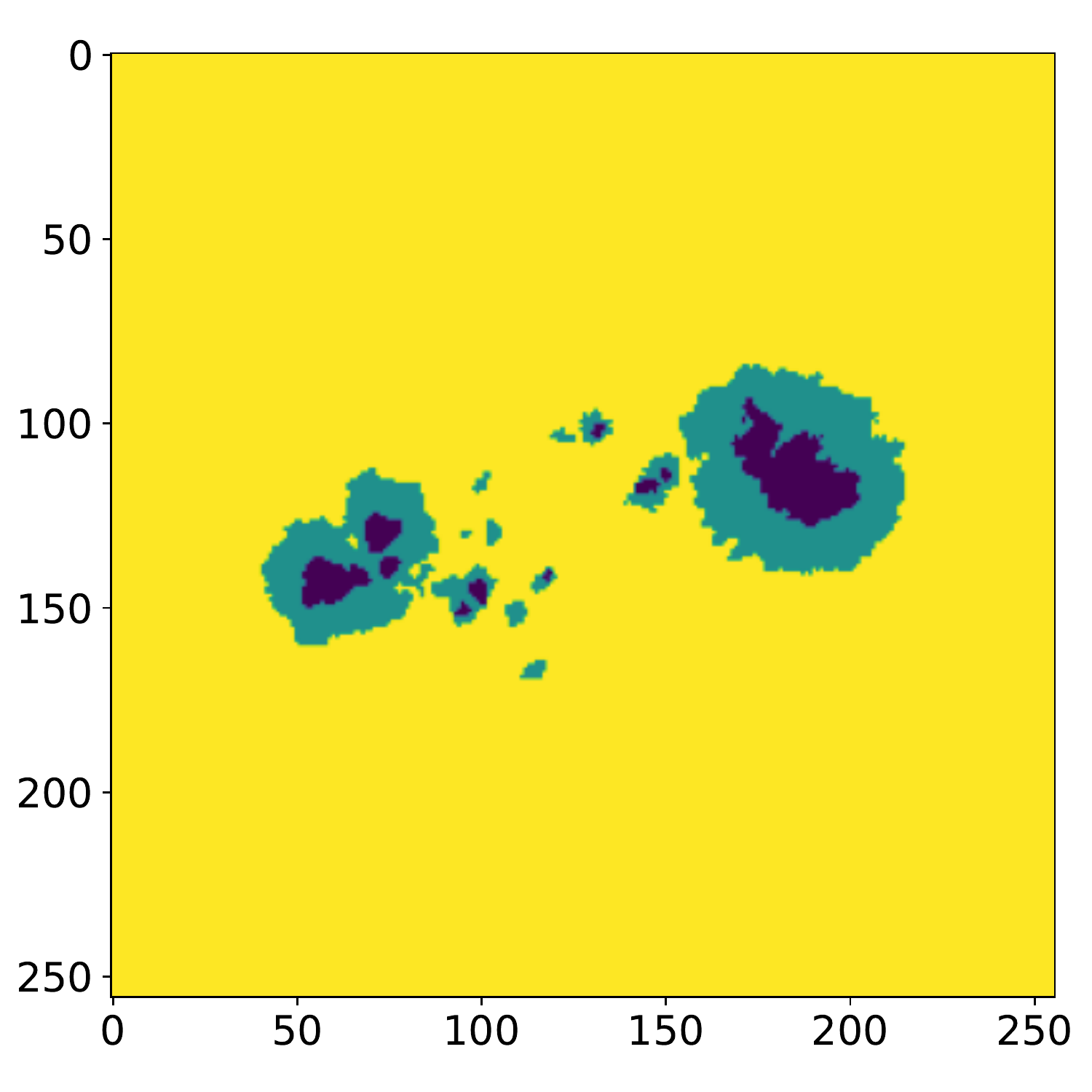}
\caption{Example of a sunspot group from the dataset. Date of observation is 2016 July 18 at 04:16UT. Sunspot group number is 124 according to the catalogue of the Kislovodsk Mountain Astronomical Station.}
\label{fig:group}
\end{figure}

For further research it was useful to combine 3 binary masks into one patch, where pixels have one of three values, 0, 1, or 2, with 0 representing photosphere (background),  1 representing sunspots and pores, 2 is for sunspot cores. Figure~\ref{fig:group} shows a patch with a sunspot group.

\section{Sunspot parameterization model}

The problem of object parameterization  is closely related to the problem of object representation in an appropriate lower-dimensional space. The latter problem is quite elaborated and there are many linear and non-linear approaches. Specifically, we will apply a neural network model with encoder-decoder architecture. There are several reasons that motivate this approach.

First, image data has extremely high-dimensional initial representation (an image of size $256\times 256$ pixels is a point in a space of size $256^2$). Standard linear dimensionality reduction models, e.g. Principal Component Analysis (PCA, see \citet{PCA}), usually work properly if the number of samples is compatible or exceeds the number of dimensions. Otherwise it might lead to inconsistent results (see e.g. \cite{HDPCA} for rigorous mathematical discussion). Thus we will focus on non-linear models, however, it should be noted, that the PCA model can nevertheless be useful in some applications (see \citet{Moon}).

Second, many models are able to work with data given in a vector form only. To process image data one has to stack rows of image matrices into a single row that breaks initial data representation. Much less models are able to process image data directly.
These are mostly the models based on convolutional operations (e.g. convolutional neural networks).
One of such models is the Variational Autoencoder (VAE, see \citet{VAE}).

\subsection{Variational Autoencoder}
\begin{figure}
\centering
\includegraphics[width=0.95\textwidth]{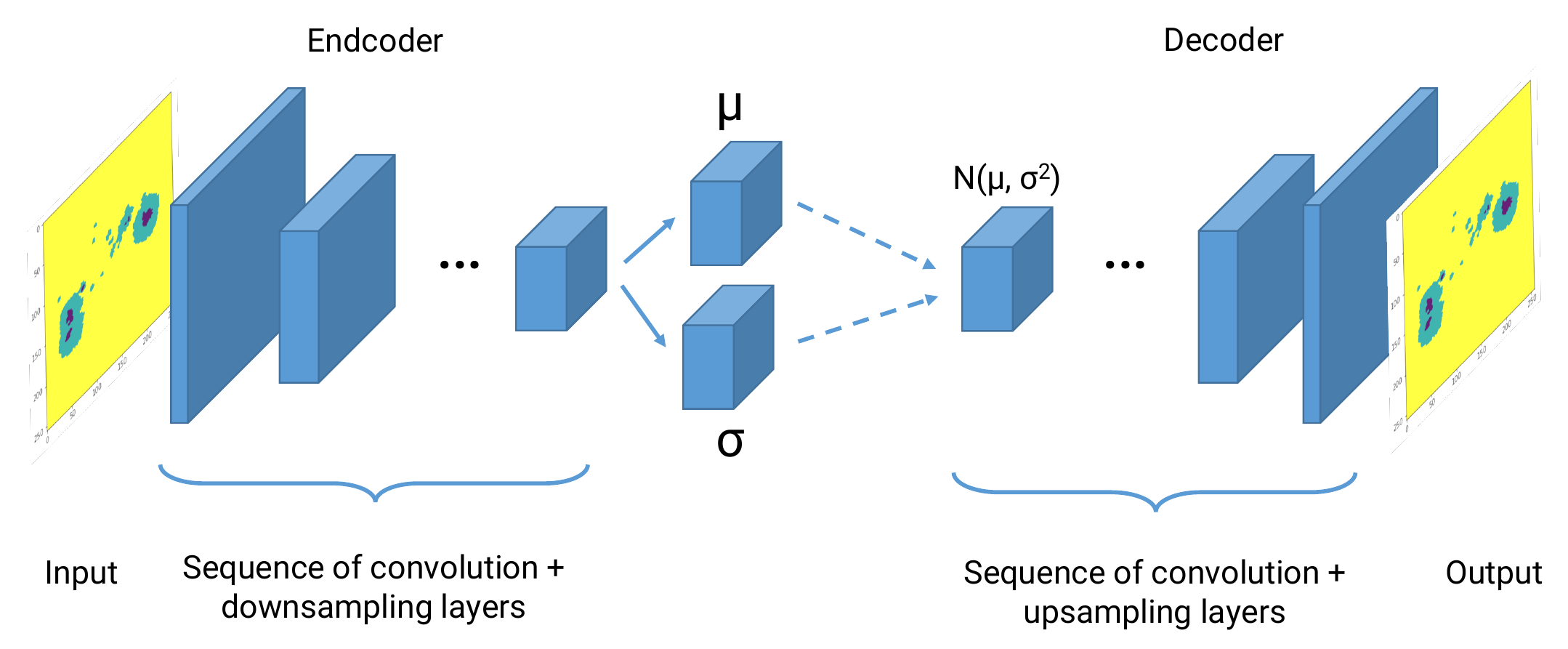}
\caption{Architecture of the Variational Autoencoder model. Encoder passes the input image through a sequence of convolution and downsampling layers. Encoder outputs two tensors, $\mu$ and $\sigma$, interpreted as vectors of mean and variance of a multivariate normal distribution. Decoder obtains a sample from that distribution and passes it through a sequence of convolution and upsampling layers. Output of the decoder is an image of the same shape as the input image in the encoder.}
\label{fig:vae}
\end{figure}

The VAE consists of two consecutive parts called an encoder and a decoder (see Figure~\ref{fig:vae}).
The encoder compresses input image into a tensor\footnote{In application to neural networks, one- or multi-dimensional arrays are called tensors.} of lower\footnote{Strictly speaking, it is not necessary that the dimensionality of the output tensor is lower that the dimensionality of input data.} size through a sequence of convolutional and downsampling layers. We interpret this tensor as a lower-dimensional representation of the input image or a \textit{latent vector}. The idea of the decoder is to reconstruct an original image from the lower dimensional representation through another sequence of convolutional and upsampling layers. Given a large dataset one can train such models to compress and decompress images without substantial loss in details.

Several additional features make the encoder-decoder model more useful.
First, we would like to obtain some regular distribution of latent vectors. 
In the framework of the Variational Autoencoder model one optimizes the model such that latent vectors converge in distribution to the multivariate standard normal distribution (standard MVN). As it is shown in Fig~\ref{fig:vae}, the encoder actually outputs two tensors, $\mu$ and $\sigma$, which we interpret as vectors on the mean and variance of the MVN. At the model training stage, the decoder actually obtains a sample from the MVN with those $\mu$ and $\sigma$ and we penalty the model using the Kullback-Leibler divergence between data distribution in the latent space and the standard MVN with zero mean and unit variance (see \citet{VAE} for more details and rigorous mathematical reasoning). Once the model is trained, we consider $\mu$ as a lower-dimensional representation of the input image and call it a latent vector.

Second, during the training stage we also penalty the model if the reconstructed image (output of the decoder) is not similar to the encoder input. To define the similarity between two images one usually applies the mean squared error (MSE) computed over pixels. However, using such a metric to train the model often leads to blurry reconstructions (see e.g. \citet{Dosovitskiy} and \citet{snell2017learning}). In our opinion this is because the MSE by definition reflects global similarity without paying enough attention to structural (local) similarity. To improve the situation we apply additionally more advanced metrics known as \textit{perceptual loss} \citep{Johnson2016Perceptual} which is based on an auxiliary pre-trained neural network model. The idea is that we measure the MSE between some intermediate layers of the pre-trained model instead of measuring the MSE between pixels of two images (specifically, we use the pre-trained VGG11 model, \citet{VGG}). In practice it allows the VAE model to generate much sharper and more natural images. We assume that it also makes the latent features more representative. 

Finer details of the VAE configuration are available in the supplementary GitHub repository along with the model training pipeline. The most essential detail is that in the bottleneck of the VAE model, which is the tensor $\mu$, we have a tensor of size 4096 (in other words, it is the dimensionality of the latent space). Given that the input image has a shape of $256\times 256$, we obtain a compression of 16 times. It does not look too compact so far, and, moreover, the latent features are not ranged in any way and some of them are quite strongly correlated.

We find that further reduction of the latent space within the VAE model complicates the training process substantially.
Instead, we find that application of the PCA model to the latent vectors $\mu$ becomes useful.

\subsection{PCA model for latent vectors}

\begin{figure}
\centering
\includegraphics[width=0.65\textwidth]{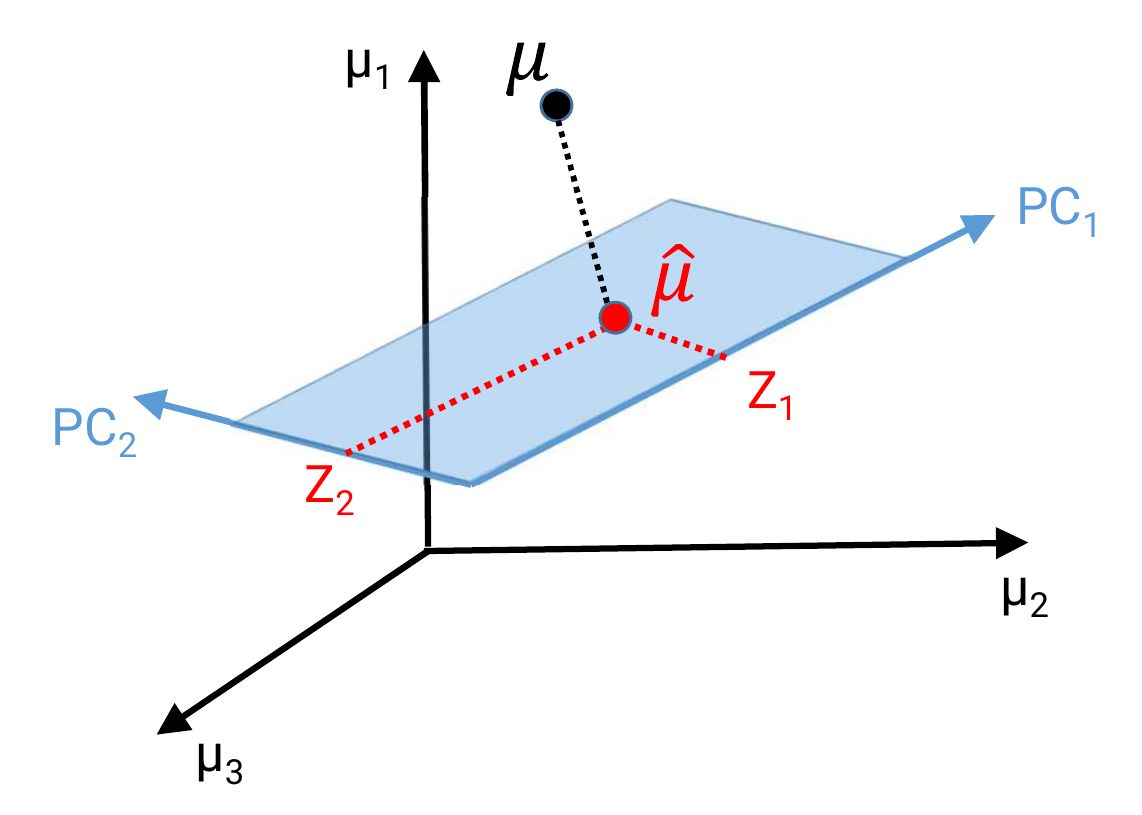}
\caption{Illustration of the PCA model. Black point $\mu$ represents a feature vector (single observation). Blue plane is the subspace formed by $n=2$ leading principal components (PCs). Red point $\hat\mu$ is the projection of $\mu$ onto the subspace formed by $PC_1$ and $PC_2$. $Z_1$ and $Z_2$ denote coordinates of $\hat\mu$ in the basis $PC_1$ and $PC_2$ and are referred to as latent vector. Squared length of the black dotted line is the reconstruction error. Given a set of feature vectors (dataset), the total reconstruction error is the sum of reconstruction errors for each observation.}
\label{fig:pca}
\end{figure}

Recall that the idea of the PCA model is to find an orthogonal basis of order $n$ that best represents the given dataset. It can be shown that this basis should be composed of $n$ leading (i.e. ordered according to the eigenvalues) eigenvectors computed from the correlation matrix of the observed variables. The eigenvectors ordered according to the eigenvalues are called \textit{principal components} (PCs). The PCA model guarantees that the obtained basis minimizes the total reconstruction error and it is also the subspace with maximal data variance (see e.g. \citet{Murphy} for mathematical details and Figure~\ref{fig:pca} for notations).

\begin{figure}
\centering
\includegraphics[width=0.75\textwidth]{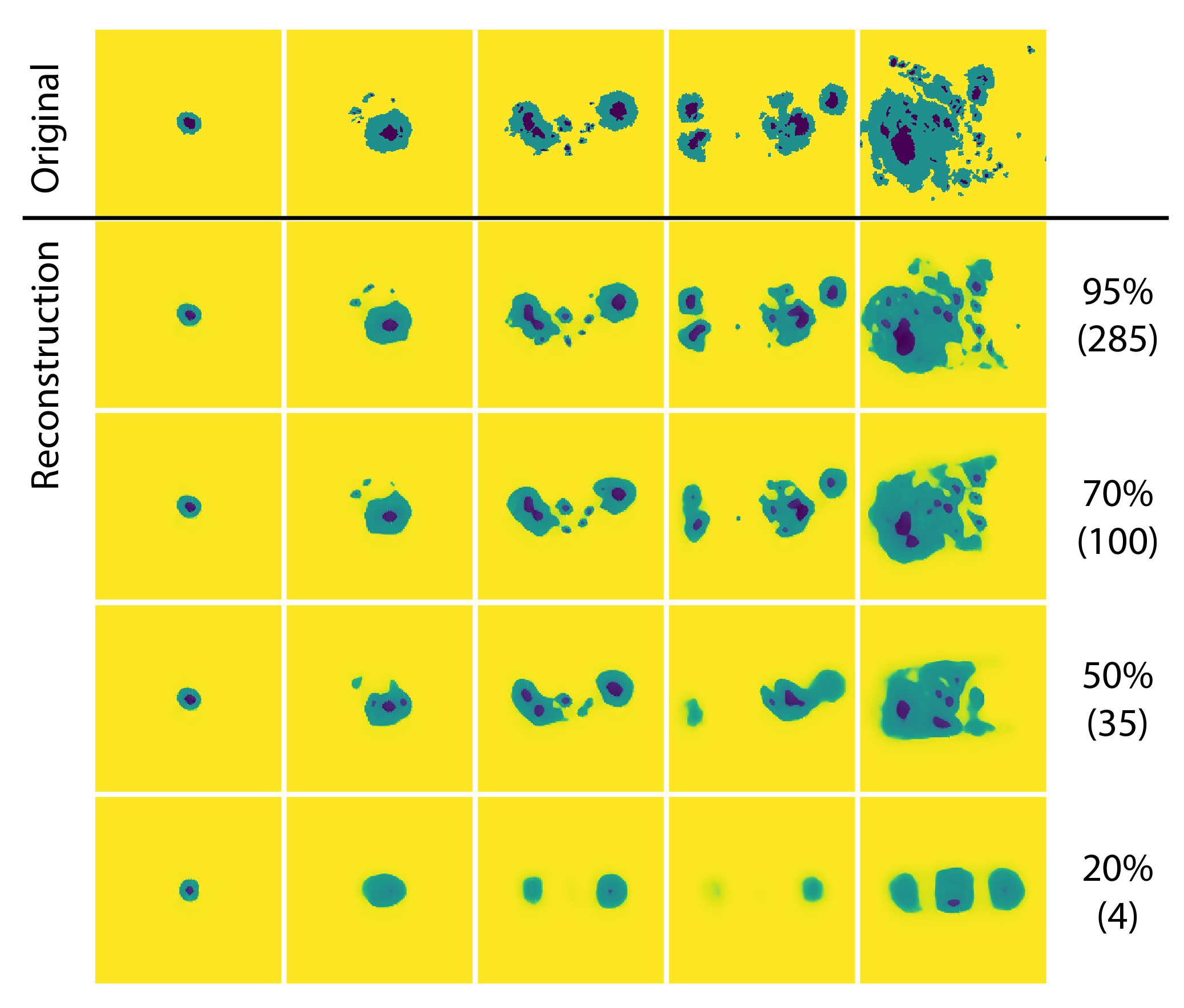}
\caption{Reconstruction accuracy against the number of principal components in the latent space of the VAE encoder. The top row shows sample sunspot groups. The next rows show reconstruction from various numbers of principal components. Percentages near each row
show explained variance, while the numbers below indicate the number of principal components. For example, 285 principal components explain 95\% of data variance.}
\label{fig:ratio}
\end{figure}

To apply the PCA model we first derive the latent vector $\mu$ (of size 4096) for each sunspot group using the VAE encoder. Then we consider the latent vectors $\mu$ as input vectors to the PCA model and derive principal components. 
In Figure~\ref{fig:ratio} we show how the number $n$ of principal components affects the accuracy of the reconstruction. We find that we need $n=285$ principal components to get 95\% of the initial variation. We will denote $Z_1$, $Z_2$, ..., $Z_{285}$ coordinates of the latent vector $\mu$ in the basis of principal components.

It can be noted in Figure~\ref{fig:ratio} that simpler groups (e.g. single-spot groups) require fewer principal components (fewer $n$) for accurate reconstruction. We will use this observation to estimate complexity of sunspot group structure (see the Appendix).

\subsection{The joint VAE and PCA model}

Usually, outputs of the VAE encoder and PCA models are referred to as latent vectors. To avoid confusion, we will denote output of the VAE encoder as latent vector $\mu$ and output of the PCA model as latent vector $Z$. Thus, the joint parametrization model works as follows: Image $\to$ VAE encoder $\to$ Latent vector $\mu$ $\to$ PCA $\to$ Latent vector $Z$. Latent vector $Z$ has size 285 and its components are ordered according to their importance estimated by the PCA model. Given an arbitrary point in the latent space $Z$ (i.e. a vector of size 285) we can reconstruct the corresponding image as follows: Latent vector $Z$ $\to$ Inverse PCA $\to$ Latent vector $\mu$ $\to$ VAE decoder $\to$ Image. In the next section we will give a physical interpretation of the latent parameters $Z$ learnt from the dataset of sunspot group images.

\section{Interpretation of latent features}

Let $Z_1$, $Z_2$, ..., $Z_{285}$ are components of the latent vector $Z$. To get an interpretation of a particular component we will set all components to zero except one and vary the remaining one and analyze the decoded images.

\begin{figure}
\centering
\includegraphics[width=0.85\textwidth]{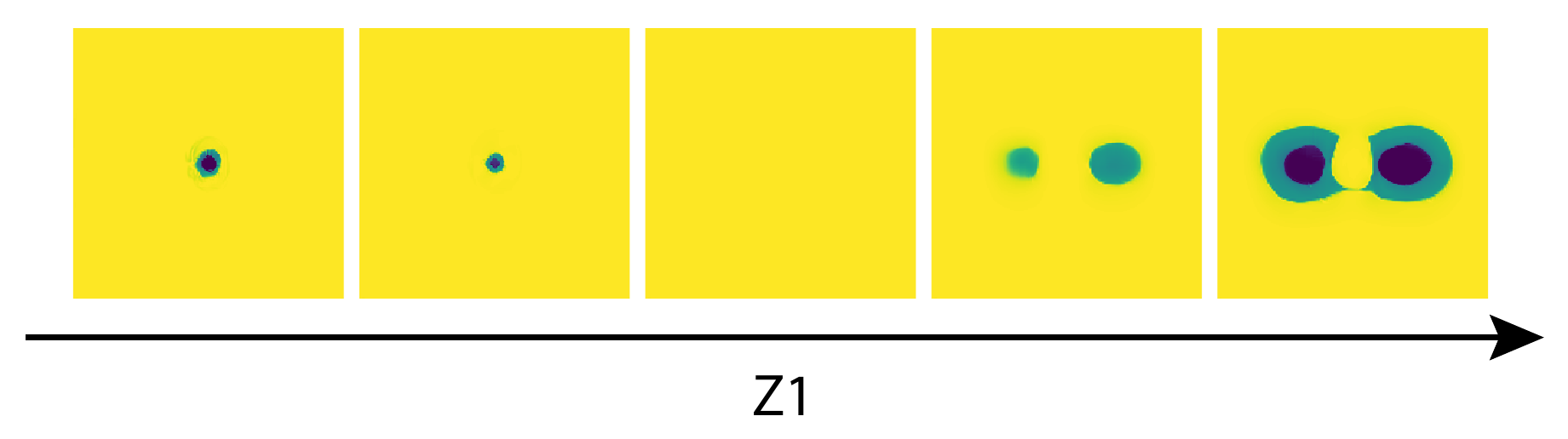}
\caption{Generated sunspot group images for latent feature $Z_1$ values from $-20$ to $20$ with step 10.}
\label{fig:z1}
\end{figure}

We start with $Z_1$. Figure~\ref{fig:z1} shows what happens with the decoded images when we vary $Z_1$ from -20 to 20 with step 10 (while $Z_2=Z_3=,,,=Z_{285}=0$).
Looking at the decoded images in Figure~\ref{fig:z1} we conclude that $Z_1$ is responsible for the single-spot (unipolar) or multi-spot (bipolar) configuration of the group.
Taking into account that this physical property is one of the most important in sunspot group characterization, it is interesting to note that the proposed parametrization model arrived at the same conclusion without any help from our side. 

\begin{figure}
\centering
\includegraphics[width=0.65\textwidth]{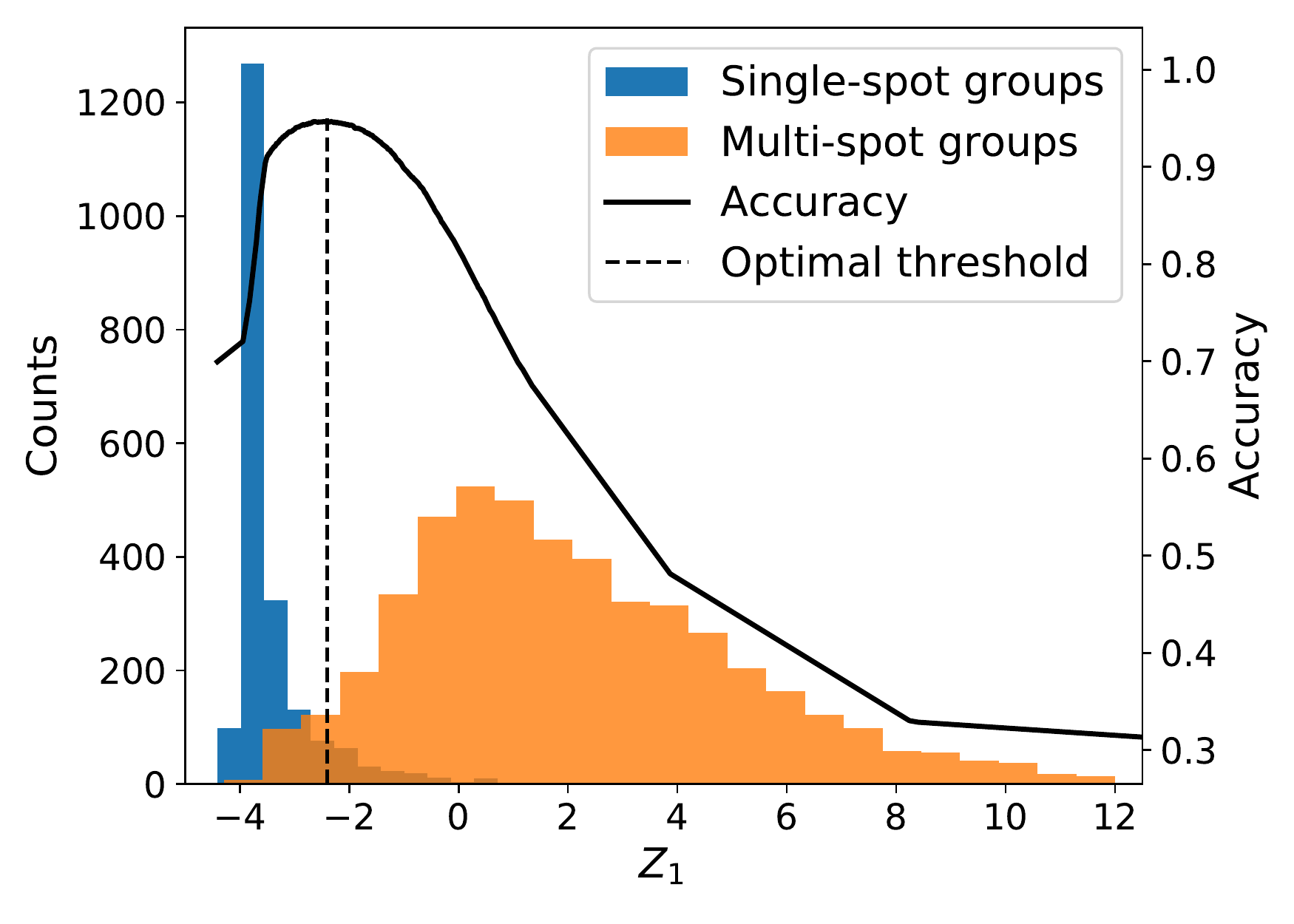}
\caption{Color histograms show distribution of the latent parameter $Z_1$ over single-spot groups (blue color) and multi-spot groups (orange color). Solid line shows accuracy of the thresholding classifier while dashed line gives the thresholding value with the highest accuracy.}
\label{fig:1dhist}
\end{figure}

To support the proposed interpretation we derive $Z_1$ for all sunspot groups in the dataset and plot the distribution in Figure~\ref{fig:1dhist}. 
Specifically, we first plot the distribution of $Z_1$ over all single-spot groups and then over all multi-spot groups. As a result we observe the two-mode distribution, where the first mode mostly corresponds to single-spot groups, and the second mode mostly corresponds to multi-spot groups. To quantify confusion statistics, we build a set of thresholding classifiers based on different threshold values $Z_1$. The classifier attributes a sunspot group to the class of single-spot groups if $Z_1$ is less than the threshold and to the class of multi-spot groups otherwise. Solid line in the Figure~\ref{fig:1dhist} shows accuracy of such classifiers against various threshold values $Z_1$. We find that the highest accuracy is 0.95 for $Z_1=-2.4$.

\begin{figure}
\centering
\includegraphics[width=0.55\textwidth]{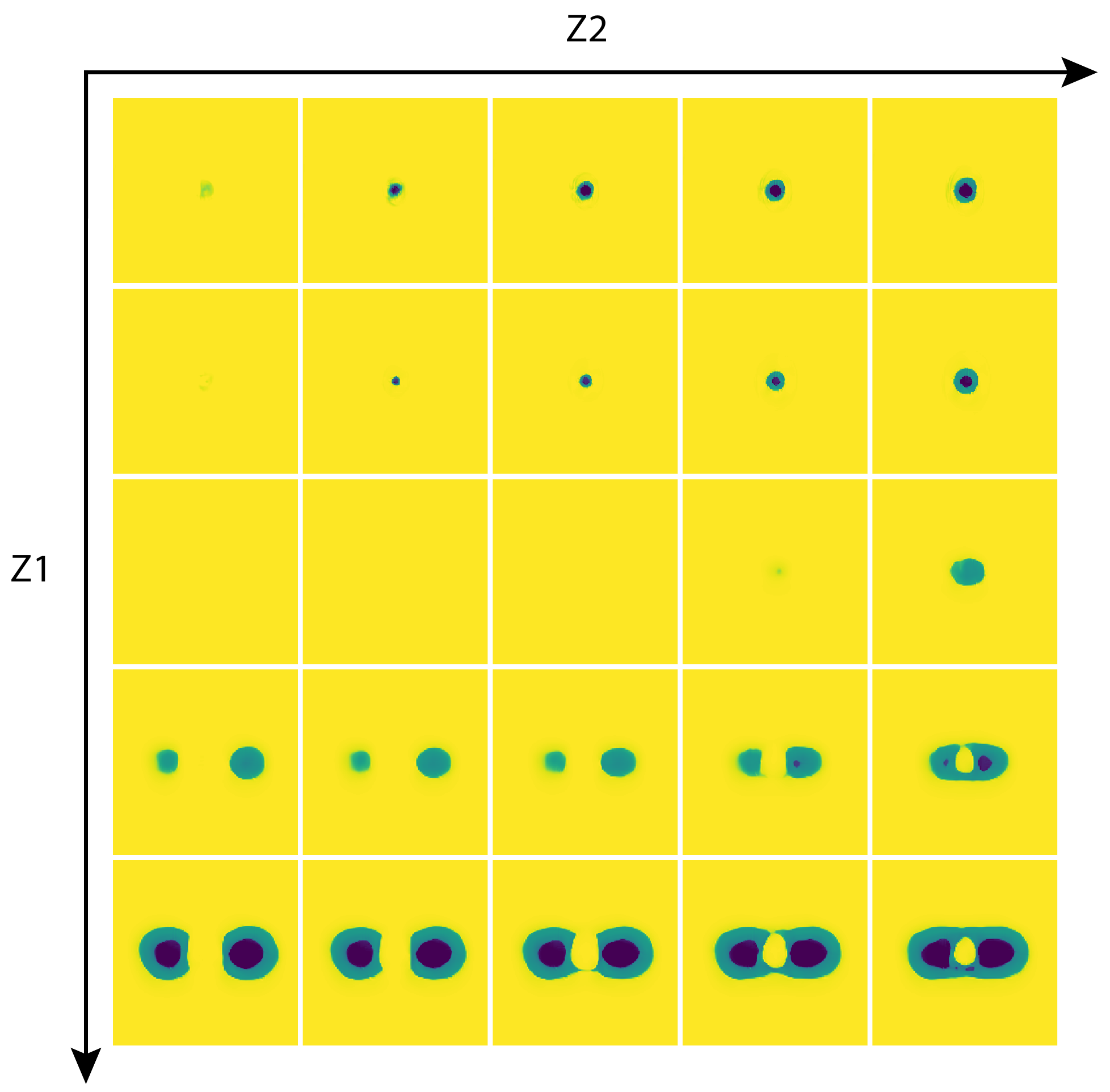}
\caption{Generated sunspot group images for various values of latent features $Z_1$ and $Z_2$. Here we vary $Z_1$ from $-20$ to $20$ step 10 and 
$Z_2$ from $-5$ to $5$ step 2.5.}
\label{fig:z1z2}
\end{figure}

\begin{figure}
\centering
\includegraphics[width=0.55\textwidth]{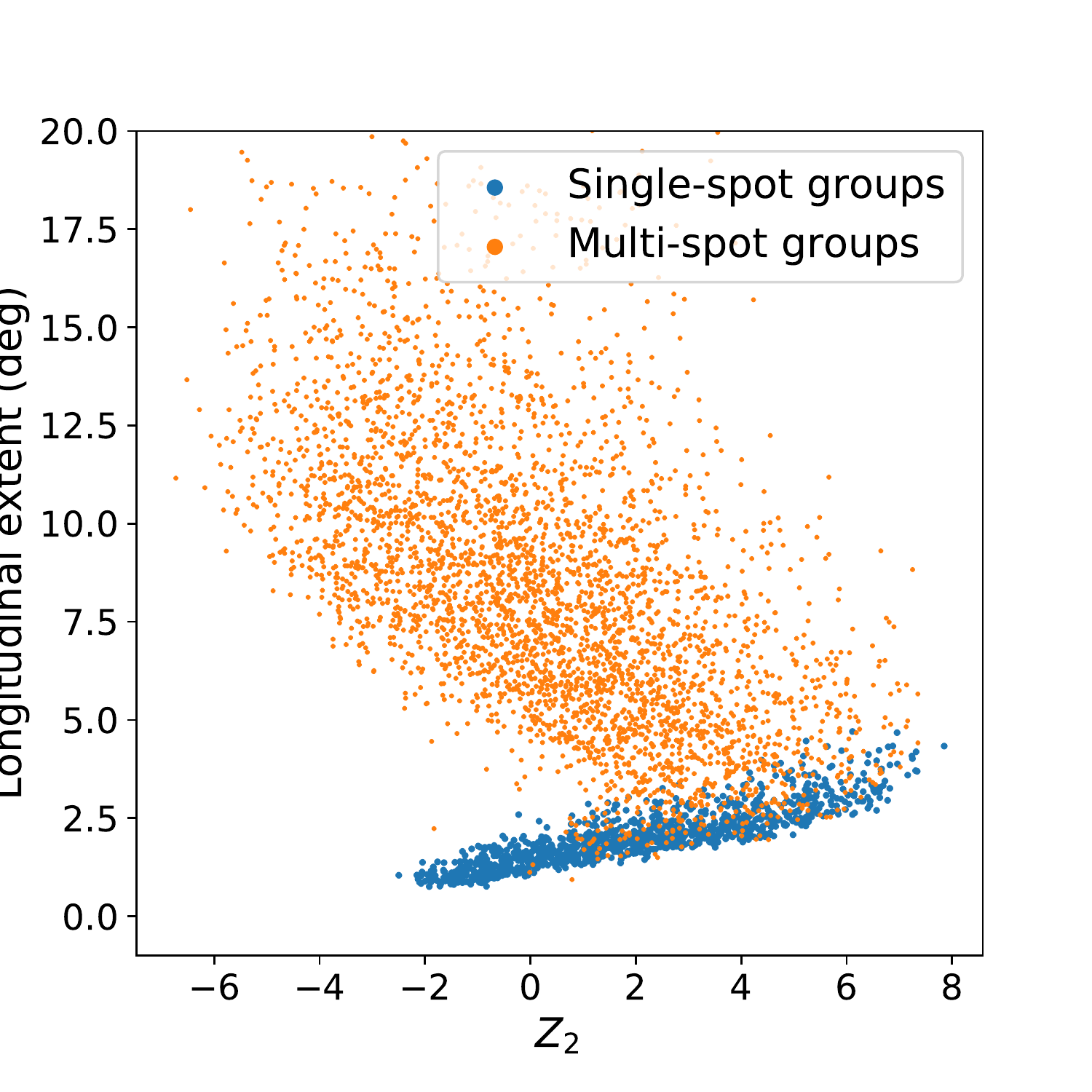}
\caption{Correlation between latent feature $Z_2$ and longitudinal extent of sunspot groups. Blue color is for single-spot groups, orange is for multi-spot groups.}
\label{fig:z2}
\end{figure}

In Figure~\ref{fig:z1z2} we investigate features $Z_1$, $Z_2$ together (i.e. we vary both $Z_1$ and $Z_2$ and plot the decoded images).
We conclude that $Z_2$ is responsible for the size of the group or, more specifically, its longitudinal extent. Indeed, in Figure~\ref{fig:z2} we compare $Z_2$ and longitudinal extent of the group and observe a strong positive correlation between these properties for unipolar group and 
negative correlation for bipolar groups.

\begin{figure}
\centering
\includegraphics[width=0.95\textwidth]{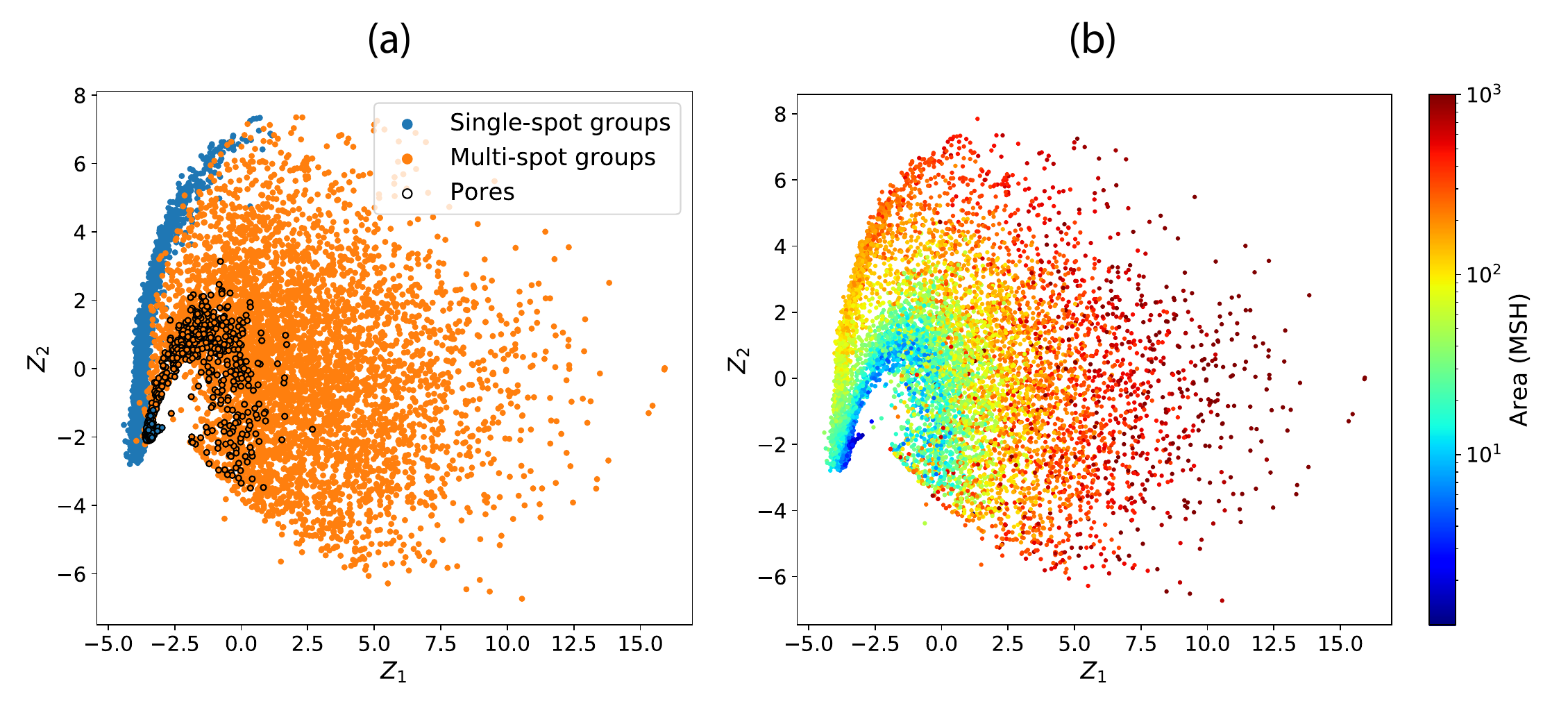}
\caption{Distribution of sunspot groups in the space of latent parameters $Z_1$ and $Z_2$. Colors in the panel (a) show single-spot and multi-spot groups. Dots with black edges mark sunspot groups represented by pores only. Colors in the panel (b) show sunspot group area measured
in millionth of solar hemisphere (MSH). }
\label{fig:2dplots}
\end{figure}

Figure~\ref{fig:2dplots} shows the distribution of sunspot groups in coordinates $Z_1$ and $Z_2$, while colors represent physical properties of the group such as number of spots and area.
First we note that the latent space has a certain structure. Although we can not explain why it looks in this particular way, we obtain that the structure is quite stable against optimization strategies used for neural network training, depth of the VAE and number of training parameters. It could be interesting to investigate separately topological properties of the manifold formed in the latent space. 

Second, in Figure~\ref{fig:2dplots}(a) we find a clear separation between single-spot and multi-spot groups. In accordance with Figure~\ref{fig:z1} the separation is mostly explained by the value of the latent parameter $Z_1$. We also find that sunspot groups represented by pores are also localized in the latent space, however, they can not be isolated based on $Z_1$ and $Z_2$ only.

Figure~\ref{fig:2dplots}(b) shows that using $Z_1$ and $Z_2$ one can estimate an area of the sunspot group.

\begin{figure}
\centering
\includegraphics[width=0.55\textwidth]{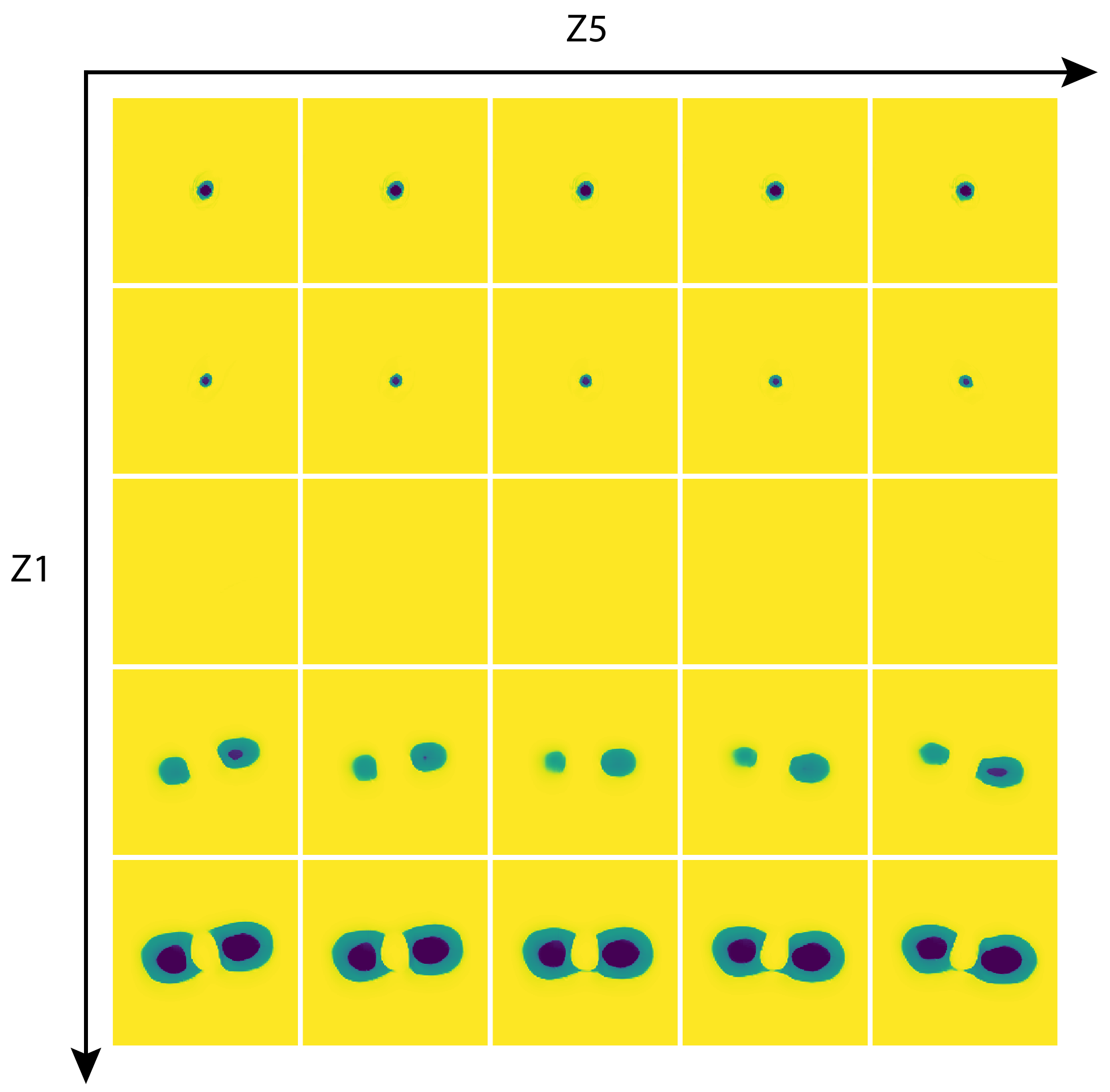}
\caption{Generated sunspot group images for various values of latent features $Z_1$ and $Z_5$. Here we vary $Z_1$ from $-20$ to $20$ with step 10 and 
$Z_5$ from $-5$ to $5$ with step 2.5.}
\label{fig:z1z3}
\end{figure}

Investigating further latent parameters, namely, $Z_3$ and $Z_4$, we find them similar up to some extent to $Z_2$, and do not discuss them. However, we find a remarkable role of $Z_5$. Figure~\ref{fig:z1z3} shows decoded images obtained for various values of $Z_1$ and $Z_5$ and we note that $Z_5$ defines the inclination (tilt) angle of the bipolar group. 
 
\begin{figure}
\centering
\includegraphics[width=0.85\textwidth]{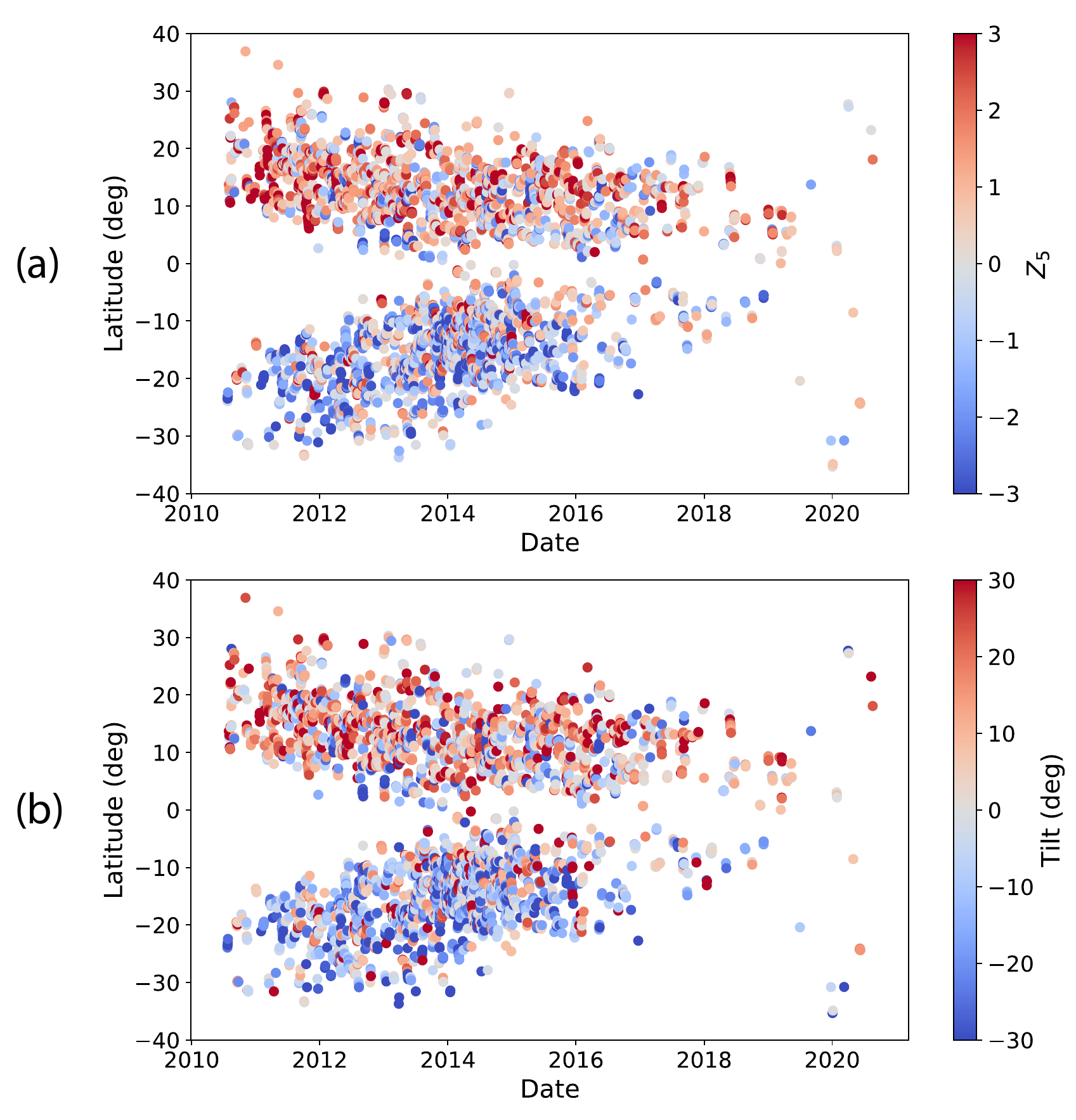}
\caption{Upper panel (a): time-latitude distribution of the latent parameter $Z_5$. Bottom panel (b): time-latitude distribution of the tilt angle measured by fitting the ordinary linear regression to the sunspot group.}
\label{fig:tilt}
\end{figure}

 The butterfly diagrams in Figure~\ref{fig:tilt} demonstrate distribution of $Z_5$ over sunspot groups as well as distribution of tilt angle measured by ordinary linear regression fitted to the multi-sunspot group \citep{Illarionov15}. The opposite signs in the northern and southern hemispheres expected for the tilt angle and observed in Figure~\ref{fig:tilt}(b) are also reproduced by $Z_5$ in Figure~\ref{fig:tilt}(a).

\begin{figure}
\centering
\includegraphics[width=0.45\textwidth]{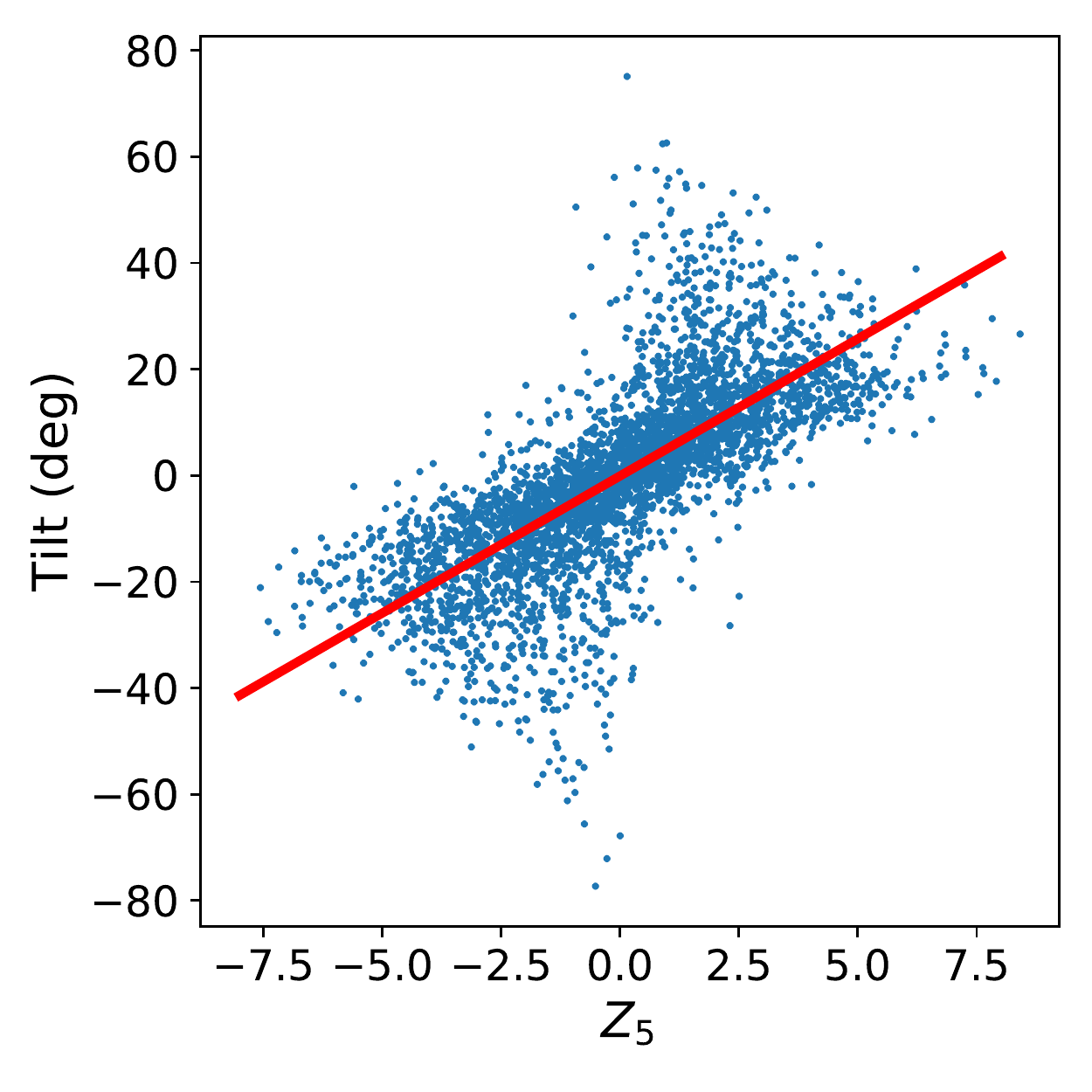}
\caption{Correlation of the latent parameter $Z_5$ and tilt angle measured by fitting the ordinary linear regression to the sunspot group. Slope of the regression line is found to be 5.2. }
\label{fig:reg}
\end{figure}

Using the slope parameter of the regression line shown in Figure~\ref{fig:reg} we estimate that tilt angle (in degrees) can be approximated by $5.2Z_5$.

\begin{figure}
\centering
\includegraphics[width=0.65\textwidth]{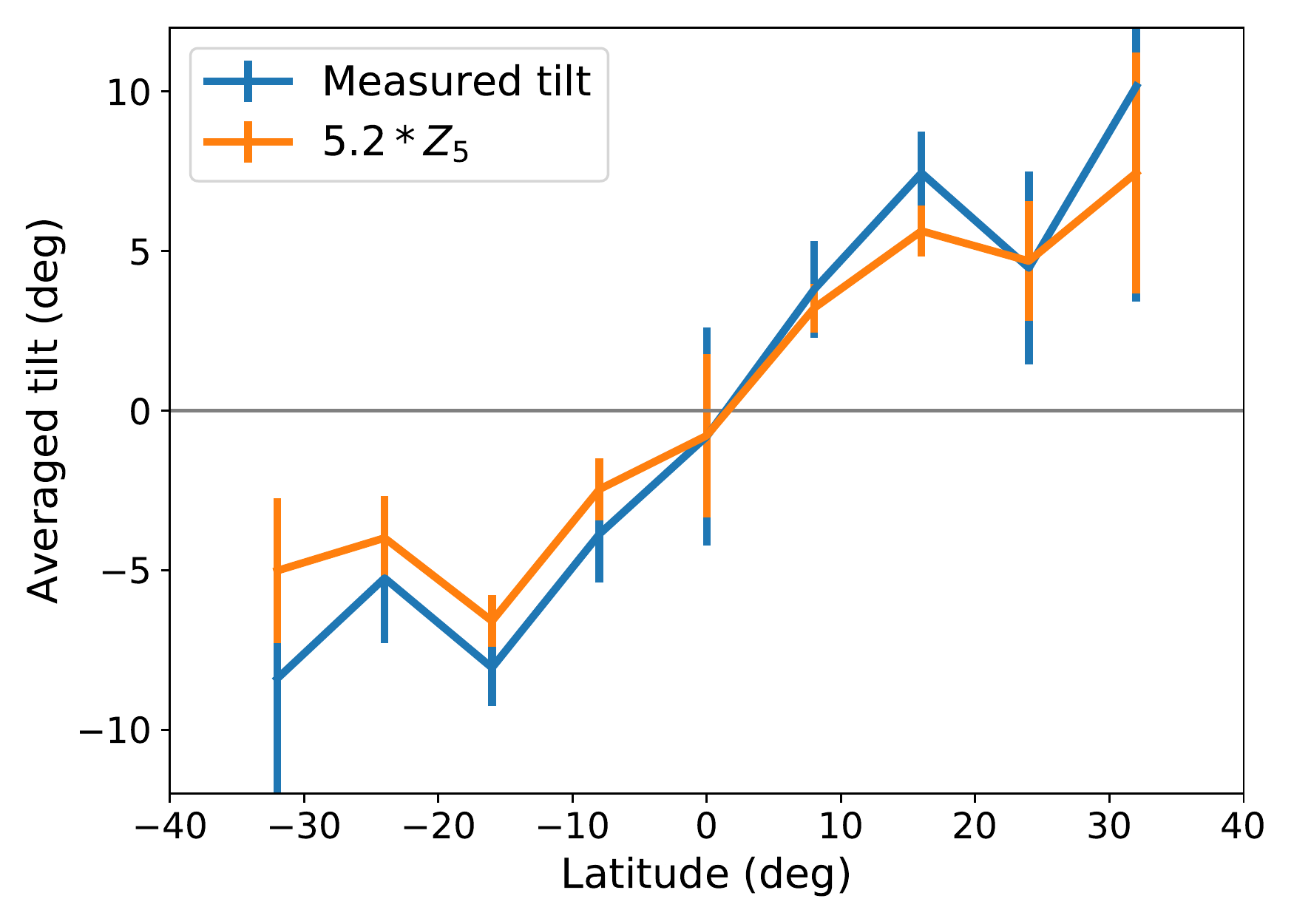}
\caption{Averaged tilt angle against latitude. Blue line is according to tilt angles measured by the regression line fitted into the sunspot group. Orange line shows tilt angles approximated by the latent parameter $Z_5$.}
\label{fig:joy}
\end{figure}

Finally, in Figure~\ref{fig:joy} we demonstrate the dependence of mean tilt angle from latitude computed
from explicitly measured tilt angles and derived from $Z_5$.
The latitude dependence observed in Figure~\ref{fig:joy} is known as Joy's law \citep{Hale}. It is interesting to note that using the latent parameter $Z_5$ one could obtain this law without measuring the tilt angle explicitly and moreover even without the concept of tilt angle. In our opinion this is an instructive example of learning useful relationships directly from the data. 

Investigating the next parameters $Z_6$, $Z_7$, ... we also observe that the role of some of them can be explained, however, the explanation becomes more complex as just polarity, area or tilt angle. For example, $Z_9$ defines the ratio between the sizes of spots in bipolar groups (see the supplementary GitHub repository).

The observed correlation between individual latent parameters and sunspot group properties can be easily improved by using a set of latent parameters and an auxiliary  trainable model. Moreover, the importance of specific latent parameters for the reconstruction of sunspot group properties can be investigated.

\begin{figure}
\centering
\includegraphics[width=0.65\textwidth]{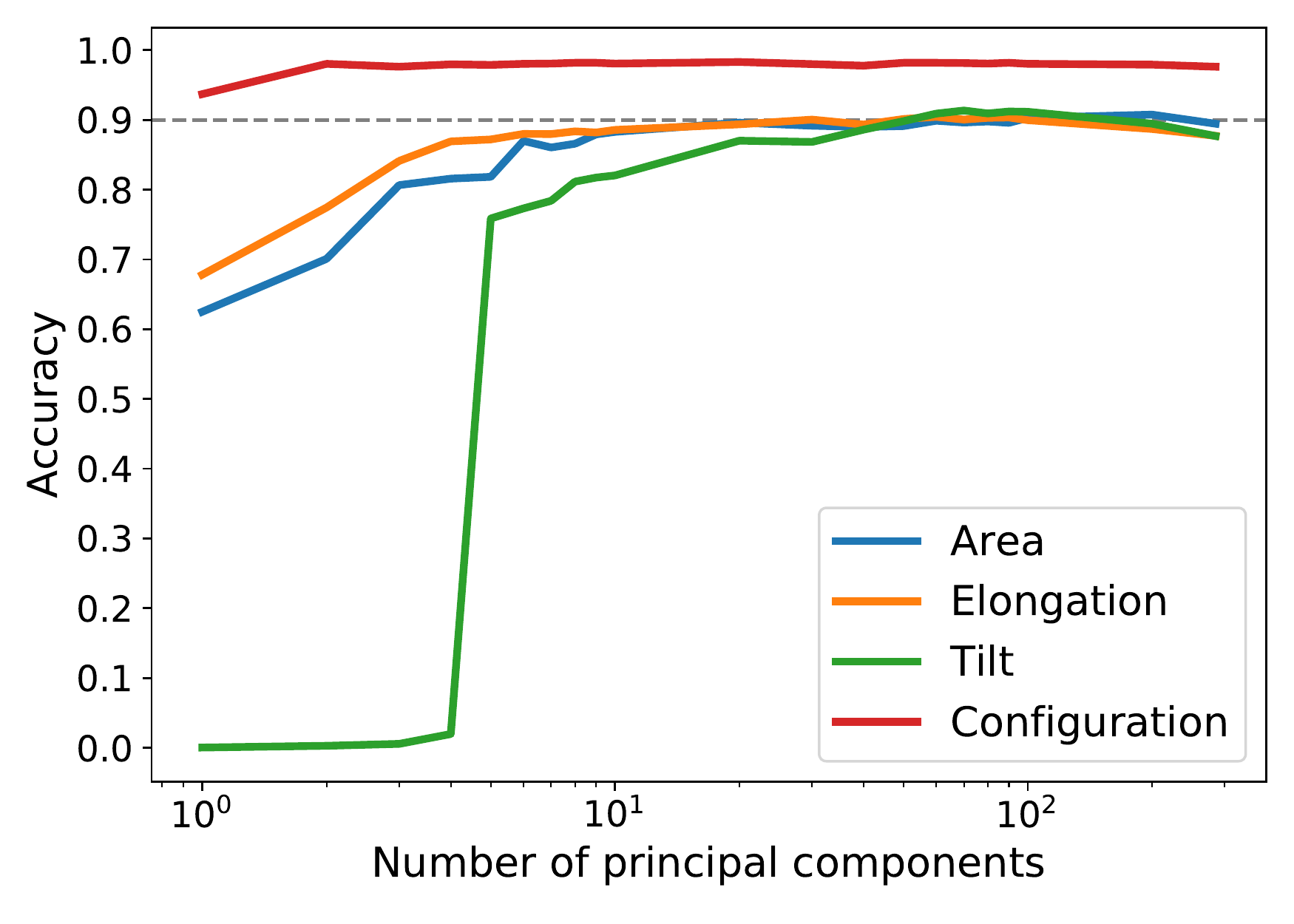}
\caption{Accuracy of reconstruction of sunspot group properties against the number of principal components (latent parameters) used. For the regression problems (estimation of the area, elongation and tilt) accuracy is defined as the $R^2$ score (coefficient of determination). For the classification problem (single- vs multi-spot) we evaluate the accuracy score (ratio of correctly classified samples).}
\label{fig:prediction}
\end{figure}

In more detail, we consider a set of the first $n$ principal components, i.e. $Z_1$, $Z_2$, ..., $Z_n$, and for each $n$ from 1 to 285 we train a simple fully-connected neural network model\footnote{The model consists of 3 hidden layers with 128, 64 and 32 neurons with ELU activation function. The output layer has a single neuron with the linear activation. We use the MSE loss function for regression problems and binary cross-entropy for the classification problem.} to map these components into some sunspot group property (area, elongation, tilt angle and configuration, which we define here as binary single- vs multi-spot classification problem). To evaluate the trained models, we use the validation set of sunspot groups (which is 30\% of the total dataset size) and compute the $R^2$ score (coefficient of determination) for regression problems (area, elongation, tilt) and accuracy score for the classification problem (single- vs multi-spot). The results are shown in Figure~\ref{fig:prediction} (note that the best possible score is 1.0 both for $R^2$ and accuracy score). From this figure, we can conclude that the first 10 latent components provide determination of the key sunspot group properties with accuracy above 0.8, while the first 40 latent components provide accuracy above 0.9.

Thus we conclude that the standard sunspot group descriptors are embedded in latent parameters and the full latent vector $Z$ of size 285 can be considered as an extended sunspot group description. The ability to reconstruct an initial image from the latent vector shows that this description is almost complete. 

In Appendix we demonstrate an application of latent vectors $Z$ to estimation of complexity of sunspot group structure and sunspot group classification.

\section{Discussion and conclusions}

We proposed a model for parametrization of sunspot groups observed in white-light. The model provides a set of 285 parameters that almost completely describe the apparent structure of sunspot groups. Although these parameters arise in the latent space as a result of machine-learning procedure, we find that some of these parameters have a clear physical interpretation. Specifically, we find parameters responsible for the unipolar or bipolar configuration of the sunspot group, size and rotation of the group. Thus, one can consider the obtained set of parameters as an extension of the set of standard sunspot group descriptors.

It should be noted that the model can be applied to sunspot observation obtained with any instrument since the only information required to the model is contours of sunspot umbras and penumbras. We consider processing of a large set of historical observations in future works as well as application to modern data obtained e.g. from the Solar Dynamics Observatory (SDO, \citet{SDO}). 

Now we discuss the relation between VAE and PCA models and why its joint application is an advantage. 

First, we find that applying the PCA model to image data directly one needs much more components to obtain the reconstruction compatible with what we obtain using joint VAE and PCA models. To be more detailed, direct application of PCA requires about 1K components while in our approach we were able to represent sunspot groups with vectors of size 285. 

Second, as it is expected from the theory, direct application of PCA to high-dimensional data might be inconsistent. It is manifested in the fact that applying small variations to the latent vector, we decode an almost noisy image. In contrast, using the joint VAE and PCA model, we obtain rather continuous latent space. In our opinion, continuity of the latent space is essential to obtain interpretable properties.

Thus we conclude that while the PCA model is attractive for application, it is useful to reduce the data dimensionality before applying the PCA. We use the VAE model for this purpose. 

At the same time, it is interesting to note that the described autoencoder neural network model can be considered as an extension of the ordinary PCA model. Indeed, replacing all non-linear activations in the autoencoder neural network model we reduce it to the linear model. It was shown by \citet{Hornik} that the linear neural network model provides the same latent space as the PCA model. Moreover, \citet{Bao} recently demonstrated that even the basis of principal components in the latent space can be recovered using an appropriate regularization. Our assumption is that using the approach proposed by \citet{Bao} one can exclude the PCA model from the parametrization model. 

One more observation is that additional connections between encoding and decoding parts of the autoencoder neural network model (so-called \textit{skip-connections}) produce another useful neural network model, called U-net. This model is widely used for image segmentation and, for example, has been applied for coronal holes detection in solar disk images and synoptic maps \citep{Illarionov_2020}.

We also find that reducing the VAE model to the standard autoencoder (AE) neural network (in other words, setting the tensor $\sigma$ from Figure~\ref{fig:vae} to zero) leads to an increase in the size of the latent vector. By training the AE model and applying the PCA model afterwards, we obtain that about 750 features are required to explain 95\% of the total variance. In contrast, the VAE+PCA model requires about 285 features to explain the same amount of total variance. Thus, VAE+PCA provides a more compact latent representation than AE+PCA.

We see at least several interesting applications of the proposed parametrization model. First, as we demonstrated in Appendix, one can use the parametrization model to estimate the complexity of sunspot groups. As a next step, it looks natural to investigate a correlation of the latent features and/or measured sunspot group complexity with other solar events, e.g. solar flares.

One more application is to develop a sunspot classification system based on latent parameters and compare it with standard ones, e.g. with the Zurich or Modified Zurich sunspot classification systems. This might reveal to what extent the standard classification systems are conditioned on data distributions (in other words, to what extent they are explained by the data distributions). The baseline results provided in the Appendix of this research can be helpful in the elaboration of automatic sunspot classification systems.

Finally, the proposed approach looks quite generic and without modifications can be applied to investigation of other traces of solar activity, e.g. prominences. It is also possible to include additional spectrum lines into consideration, e.g. to add one more channel to the input image representing the magnetic field map.  Thus it becomes possible to elaborate a data-driven magneto-morphological classification of sunspot groups as well as data-driven classification of other traces of solar activity.  

In order to propagate further research, the source code for the model training as well as the dataset of sunspot groups and obtained latent vectors are available in the public GitHub repository \url{https://github.com/observethesun/sunspot_groups}.

\begin{acks}
 The authors are grateful to the reviewers for valuable comments and suggestions. EI acknowledges the support of RSF grant 20-72-00106 and Lomonosov-2 supercomputer center at MSU for computing resources. 
\end{acks}

\appendix

\section*{Complexity of sunspot groups}

Here we elaborate an application of the parametrization model to estimation of sunspot group complexity. The main idea is that more complex structures should require more components of the latent vector $Z$ for accurate reconstruction. We convert this idea into the following procedure. First, we consider the latent vector $\mu$ in the output of the VAE encoder (recall, it has size 4096). Then we measure a distance between this vector and its projection onto the first principal component (PC) and will refer to it as \textit{initial reconstruction error}. Then we measure the reconstruction error  given the basis of the first two PCs, the first three PCs and so on. Clearly, by increasing the number of PCs, the reconstruction error will decrease. 

\begin{figure}
\centering
\includegraphics[width=0.55\textwidth]{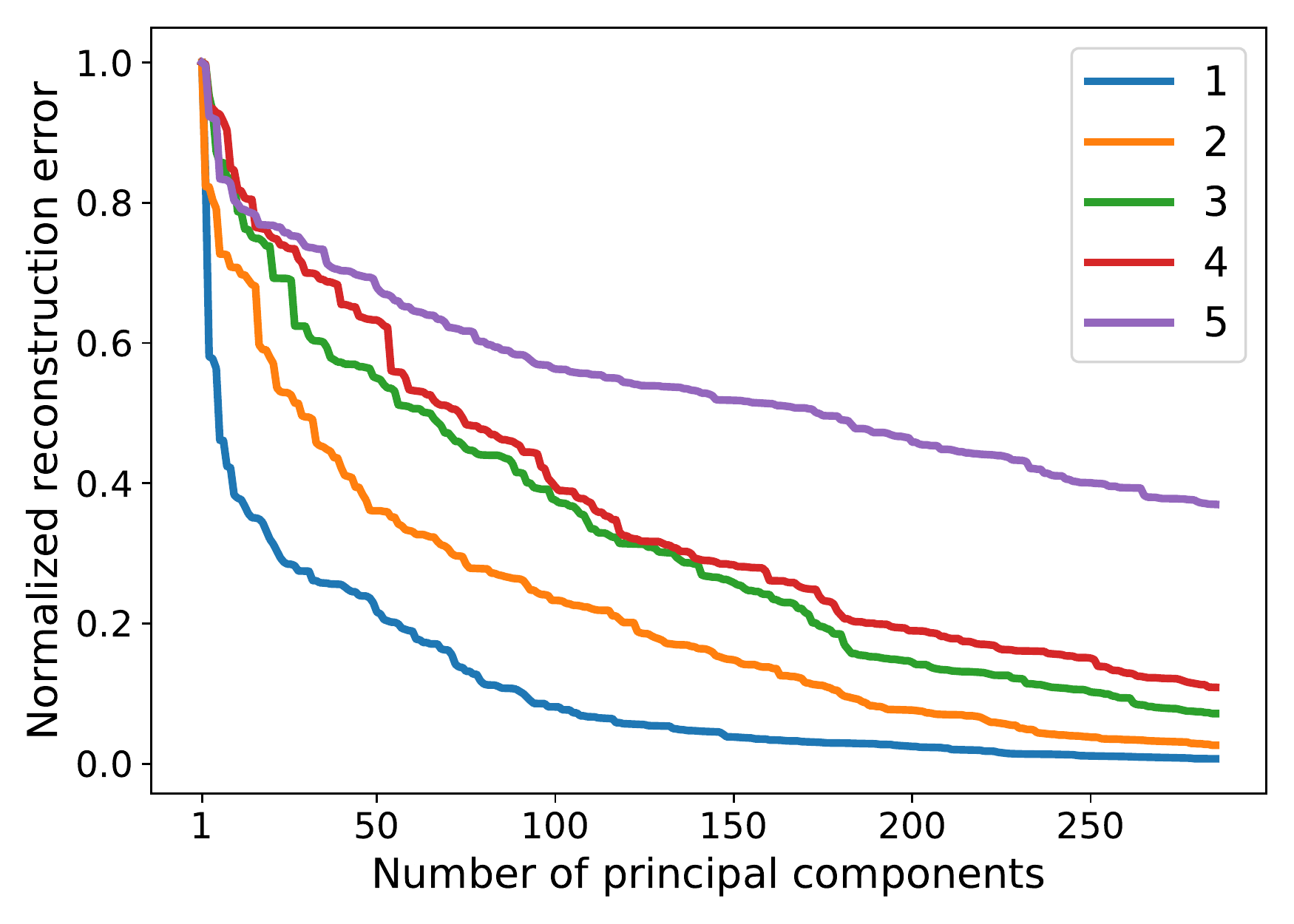}
\caption{Reconstruction error normalized to the initial reconstruction error for various number of principal components and for sunspot group images shown in the first row of Figure~\ref{fig:ratio}. Numbers in the color legend correspond to the position of the sunspot group image in the first row of Figure~\ref{fig:ratio}.}
\label{fig:errors}
\end{figure}

In Figure~\ref{fig:errors} we show different decreasing patterns that arise from increasing the number of PCs from 1 to 285. Note that colors in Figure~\ref{fig:errors} correspond to sunspot group images shown in the first row of the Figure~\ref{fig:ratio}. Intuitively, the complexity of sunspot group structures increases in the first row of the Figure~\ref{fig:ratio}. This impression is supported by the  Figure~\ref{fig:errors} where we observe that the first line drops more rapidly than the second one, the second line drops more rapidly than the third one and so on. Thus we conclude that the decreasing pattern of the reconstruction error correlates with visual estimation of sunspot group complexity.

\begin{figure}
\centering
\includegraphics[width=0.65\textwidth]{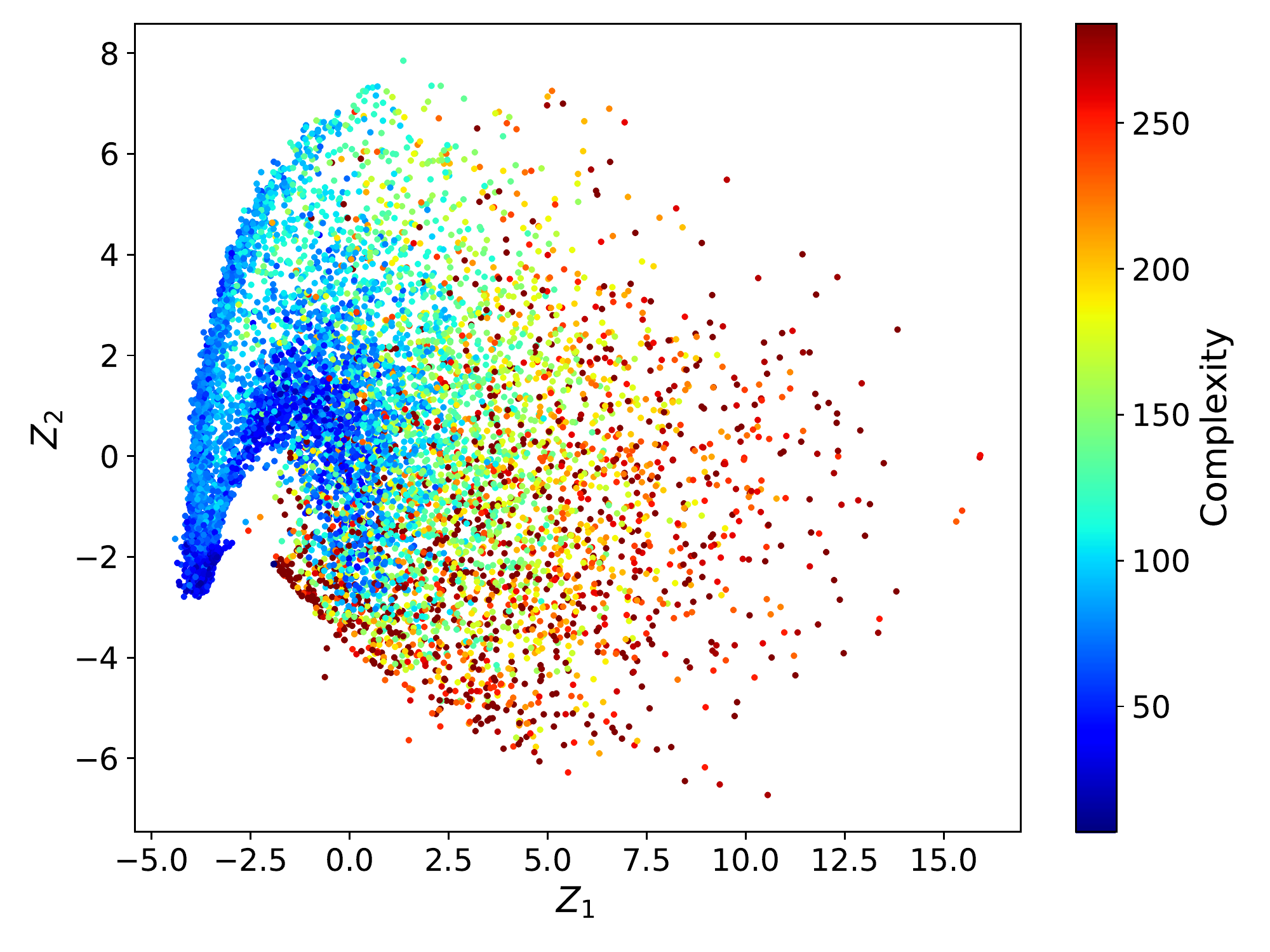}
\caption{Distribution of the complexity over all sunspot groups visualized in the space of latent parameters $Z_1$ and $Z_2$. Complexity is defined as the number of principal components at which the reconstruction error is half the initial reconstruction error.}
\label{fig:complexity}
\end{figure}

In order to quantify the complexity we will find the number of PCs at which the reconstruction error is half of the initial reconstruction error. Figure~\ref{fig:complexity} shows a distribution of the measured complexity over all sunspot groups visualized in the space of latent parameters $Z_1$ and $Z_2$.
Comparing Figure~\ref{fig:complexity} with Figure~\ref{fig:2dplots} we conclude that the proposed complexity measure corresponds to the expected properties and is low for single-spot groups and groups with small areas and increases for large multi-spot groups.

\section*{Classification of sunspot groups}

As suggested in the main text, the latent parameters can be useful for  sunspot classification. Proper investigation of this idea requires a verified annotation of sunspot groups, e.g. following the Zurich or McIntosh classification systems. The dataset we use in this research does not contain such labels.

Although there are external datasets with sunspot group classes (e.g., NOAA/USAF\footnote{\url{http://solarcyclescience.com/activeregions.html}} or Locarno\footnote{\url{https://sunspots.irsol.usi.ch/db/}} catalogues), we stress that a separate research is required to establish a proper correspondence. There are at least several reasons why this process is not trivial. First, there is a certain time lag between observations in different catalogues. Taking into account the rapid evolution of sunspot groups at early stages, this time lag can cause systematic inconsistencies. Second, the difference in the resolution of telescopes (especially, satellite and ground-based) can strongly affect the estimation of the number of small spots and the identification of sunspot cores.

\begin{figure}
\centering
\includegraphics[width=0.65\textwidth]{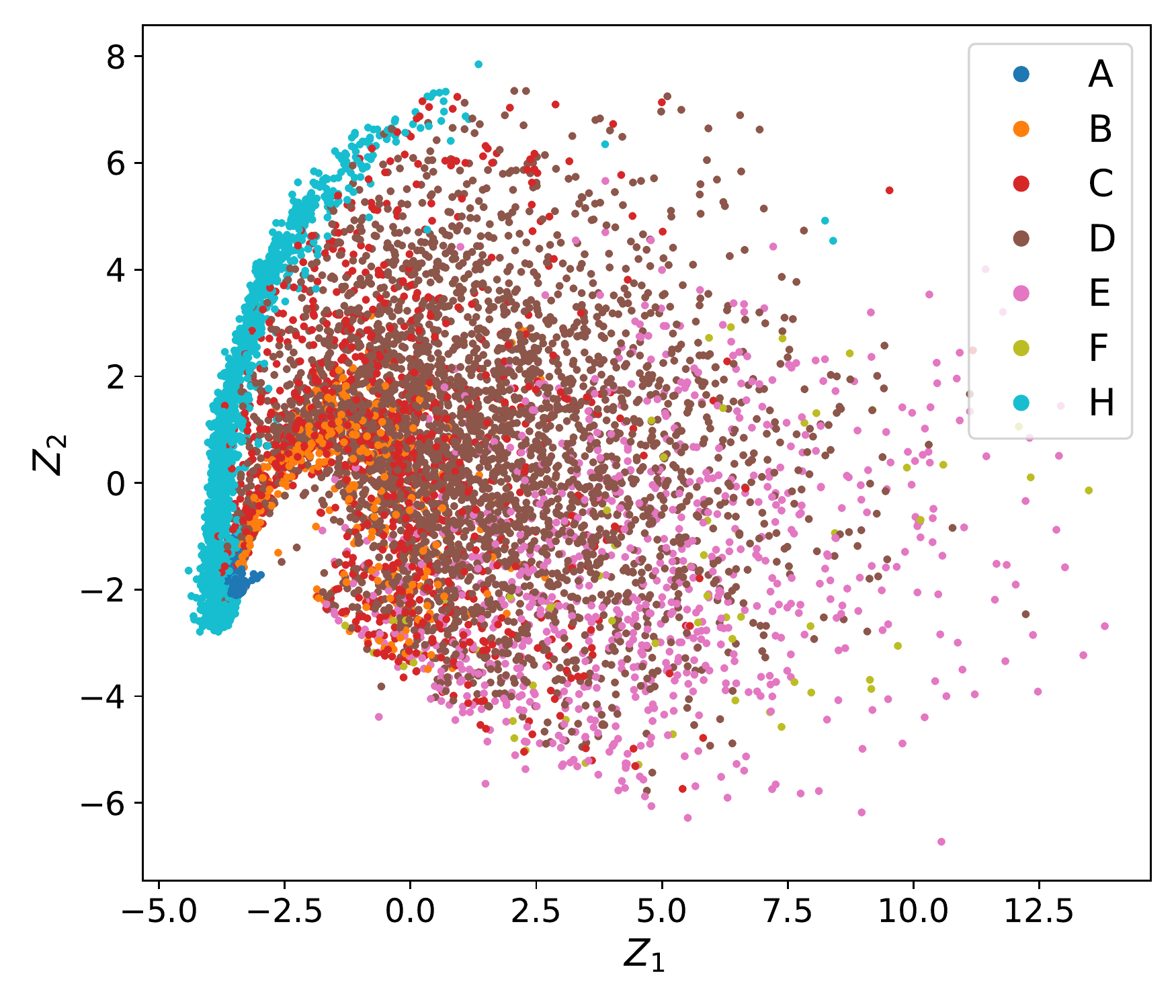}
\caption{Distribution of sunspot group labels assigned according to the Table~\ref{tab:classes} in the space of latent parameters $Z_1$ and $Z_2$.}
\label{fig:classes}
\end{figure}

Nevertheless, to demonstrate the possibility of using latent parameters to classify sunspots, in this study we introduce a synthetic classification that mimics the McIntosh one. Specifically, we assign sunspot classes according to Table~\ref{tab:classes}. Distribution of the classes in the space of latent parameters $Z_1$ and $Z_2$ is shown in Figure~\ref{fig:classes}.

\begin{table}
\caption{Algorithm for sunspot group labelling. These labels are used as targets for the classification model training.}
\label{tab:classes}
\begin{tabular}{ccccc}
\hline
    Class & Number of spots & Number of cores & Number of spots with cores & Elongation \\
\hline
    A & 1 & 0 & 0 & any\\
    B & $\ge2$ & 0 & 0 & any\\
    C & $\ge2$ & $\ge1$ & 1 & any\\
    D & $\ge2$ & $\ge2$ & $\ge 2$ & $< 10^{\circ}$\\
    E & $\ge2$ & $\ge2$ & $\ge 2$ & $< 15^{\circ}$\\
    F & $\ge2$ & $\ge2$ & $\ge 2$ & $\ge 15^{\circ}$\\
    H & 1 & $\ge1$ & 1 & any\\
\hline
\end{tabular}
\end{table}

Then we train a simple fully-connected neural network model (similar to the one previously used to estimate sunspot group properties) to predict classes based on latent vectors alone. We reserved 30\% of samples for  model validation and show the classification metrics in Table~\ref{tab:test}. We find that accuracy varies substantially between classes and is 0.75 on average.

\begin{figure}
\centering
\includegraphics[width=0.65\textwidth]{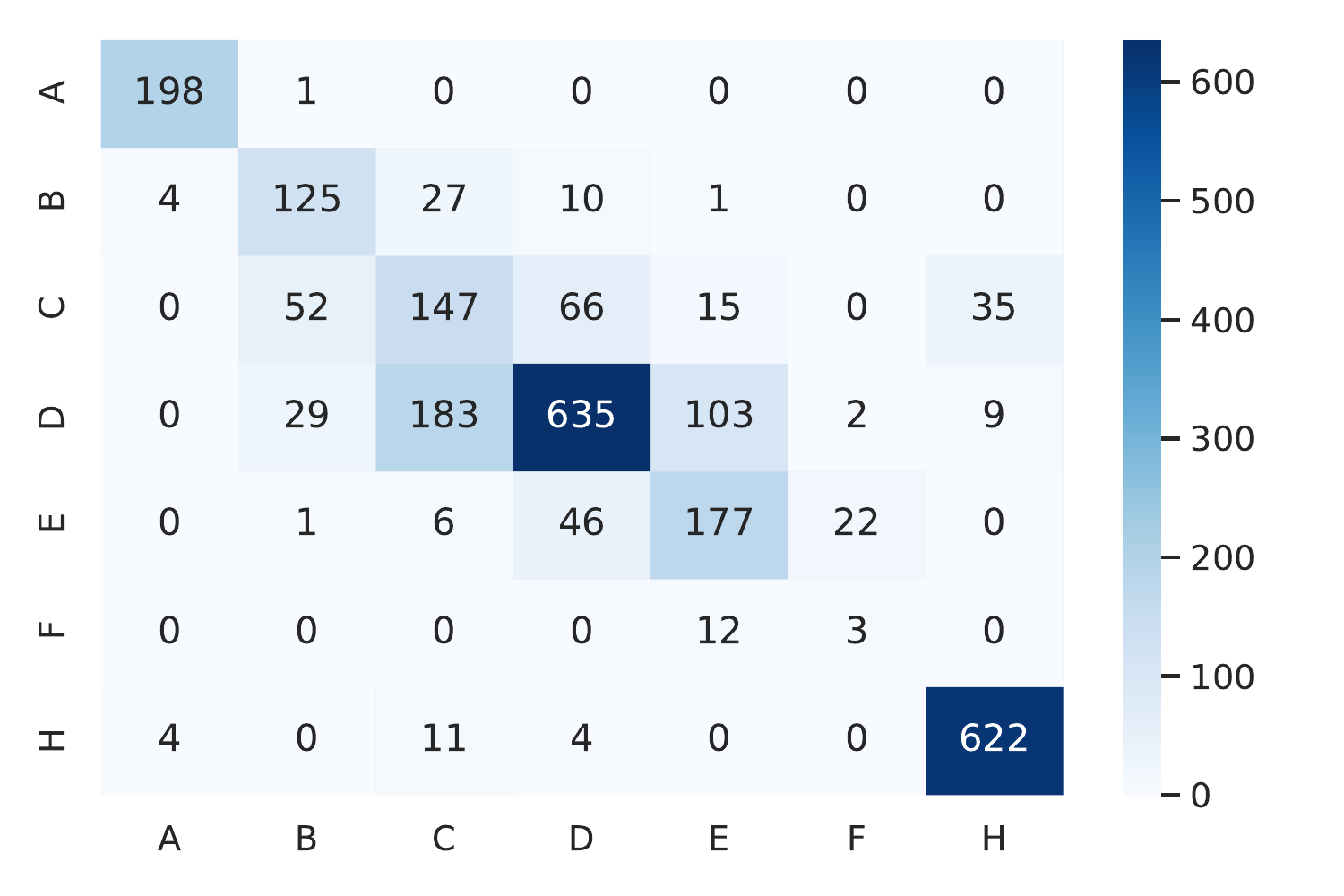}
\caption{Confusion matrix for the validation subset. True classes correspond to rows, predicted classes correspond to columns. }
\label{fig:confusion}
\end{figure}

There are several reasons that, in our opinion, limit the accuracy. First, there is a strong class imbalance in the dataset. Thus, we obtain very moderate scores for the rare classes. Second, the shallow neural network we used for the demonstration may be too simple to adequately decode latent vectors. We find that deeper models easily fail into strong overfitting. For the real application it looks reasonable to complement the latent vector with some simple sunspot group properties so that the model can benefit from both simple and deep sunspot group descriptors. Third, the confusion matrix shown in the Figure~\ref{fig:confusion} reveals that the model often confuses close classes (e.g. D and E or E and F). Given that the difference between these classes is only in the elongation of the group, the classification model can easily be improved using explicit sunspot group properties.

Finally, we would like to note that in practice sunspot group classes reflect the evolutionary stage of the group rather than the instantaneous characteristics. This means that a correct classification model should also rely on the group's prehistory. In our opinion, latent vectors can be a useful tool for studying the dynamics of sunspot groups, and we leave this study for future work.

\begin{table}
\caption{Validation metrics. Overall accuracy is 0.75.}
\label{tab:test}
\begin{tabular}{ccccc}
\hline
Class &  Precision &  Recall &  F1-score &  Support \\
\hline
A & 0.96 & 0.99 & 0.98 & 199 \\
B & 0.60 & 0.75 & 0.67 & 167 \\
C & 0.39 & 0.47 & 0.43 & 315 \\
D & 0.83 & 0.66 & 0.74 & 961 \\
E & 0.57 & 0.70 & 0.63 & 252 \\
F & 0.11 & 0.20 & 0.14 & 15 \\
H & 0.93 & 0.97 & 0.95 & 641 \\
\hline
\end{tabular}
\end{table}

\bibliographystyle{spr-mp-sola}
\bibliography{literature}

\end{article} 
\end{document}